\documentclass[lettersize,journal]{IEEEtran}
\usepackage{cite}
\usepackage{algorithmic}
\usepackage{algorithm}
\usepackage{array}
\usepackage[caption=false,font=normalsize,labelfont=sf,textfont=sf]{subfig}
\usepackage{caption}
\usepackage{textcomp}
\usepackage{stfloats}
\usepackage{url}
\usepackage{verbatim}
\usepackage{graphicx}
\usepackage[normalem]{ulem}
\usepackage{amsmath,amssymb,amsfonts}
\usepackage[dvipsnames]{xcolor}
\usepackage[acronym]{glossaries}
\makeglossaries
\newacronym{ict}{ICT}{Information and Communication Technology}
\newacronym{nwdaf}{NWDAF}{Network Data Analytics Function}
\newacronym{2d}{2D}{two-dimensional}
\newacronym{3d}{3D}{three-dimensional}
\newacronym{3gpp}{3GPP}{3rd generation partnership project}
\newacronym{4g}{4G}{fourth-generation}
\newacronym{5g}{5G}{fifth generation of mobile networks}
\newacronym{6dof}{6DoF}{six degrees of freedom}
\newacronym{6g}{6G}{sixth-generation}
\newacronym{abp}{ABP}{authentication by personalisation}
\newacronym{ac}{AC}{alternating current}
\newacronym{aclr}{ACLR}{adjacent channel leakage ratio}
\newacronym{adc}{ADC}{analog-to-digital converter}
\newacronym{adr}{ADR}{adaptive data rate}
\newacronym{aec}{AEC}{aluminum electrolytic capacitor}
\newacronym{aes}{AES}{Advanced Encryption Standard}
\newacronym{afe}{AFE}{analog front end}
\newacronym{ag}{AG}{array gain}
\newacronym{agc}{AGC}{automatic gain controller}
\newacronym{agps}{A-GPS}{Assisted Global Positioning System} %bestaat al :) % Aha, dan maak ik er wel Assisted gps van!
\newacronym{ai}{AI}{artificial intelligence}
\newacronym{alk}{ALK}{Alkaline}
\newacronym{am}{AM}{amplitude modulation}
\newacronym{amam}{AM/AM}{amplitude modulation to amplitude modulation}
\newacronym{amc}{AMC}{automatic modulation classification}
\newacronym{amp}{AMP}{approximate message passing}
\newacronym{ampm}{AM/PM}{amplitude modulation to phase modulation}
\newacronym{aoa}{AOA}{angle-of-arrival}
\newacronym{aod}{AOD}{angle-of-departure}
\newacronym{ap}{AP}{access point}
\newacronym{api}{API}{application program interface}
\newacronym{apt}{APT}{acoustic power transfer}
\newacronym{apu}{APU}{access point unit}
\newacronym{ar}{AR}{augmented reality}
\newacronym{arima}{ARIMA}{Auto-Regressive Integrated Moving Average}
\newacronym{arp}{ARP}{Antenna Reference Point}
\newacronym{asic}{ASIC}{application specific integrated circuit}
\newacronym{ask}{ASK}{amplitude-shift keying}
\newacronym{at}{AT}{ATtention}
\newacronym{auv}{AUV}{autonomous underwater vehicle}
\newacronym{awg}{AWG}{arbitrary waveform generator}
\newacronym{awgn}{AWGN}{additive white Gaussian noise}
\newacronym{baw}{BAW}{bulk acoustic wave}
\newacronym{bb}{BB}{base-band}
\newacronym{bc}{BC}{backscatter communication}
\newacronym{bd}{BD}{backscatter device}
\newacronym{be}{BE}{Belgium}
\newacronym{ber}{BER}{bit error rate}
\newacronym{bf}{BF}{beamforming}
\newacronym{bga}{BGA}{ball grid array}
\newacronym{bldc}{BLDC}{brushless DC}
\newacronym{bod}{BOD}{Brown-Out Detection}
\newacronym{bom}{BOM}{bill of materials}
\newacronym{bpsk}{BPSK}{binary phase-shift keying}
\newacronym{brp}{BRP}{beam refinement process}
\newacronym{bs}{BS}{base station}
\newacronym{bu}{BU}{booster unit}
\newacronym{bw}{BW}{bandwidth}
\newacronym{ca}{CA}{carrier emitter}
\newacronym{cad}{CAD}{channel activity detection}
\newacronym{cars}{CARS}{calibration reference signal}
\newacronym{cbm}{CBM}{condition based maintenance}
\newacronym{cc}{CC}{constant current}
\newacronym{ccdf}{CCDF}{complementary cumulative distribution function}
\newacronym{ccnn}{CCNN}{circular convolutional neural network}
\newacronym{ccs}{CCS}{correlative channel sounder}
\newacronym{cdf}{CDF}{cumulative distribution function}
\newacronym{cdma}{CDMA}{code division-multiple access}
\newacronym{cdrx}{CDRX}{connected mode DRX}
\newacronym{ce}{CE}{coverage enhancement}
\newacronym{ced}{CED}{cumulative energy density}
\newacronym{cf}{CF}{cell-free}
\newacronym{cfmmimo}{CF-mMIMO}{cell-free massive MIMO}
\newacronym{cfo}{CFO}{carrier frequency offset}
\newacronym{cir}{CIR}{channel impulse response}
\newacronym{cla}{CLA}{closed-loop approach}
\newacronym{clk}{CLK}{clock}
\newacronym{cmos}{CMOS}{complementary metal oxide semiconductor}
\newacronym{cnn}{CNN}{convolutional neural network}
\newacronym{co}{CO}{combinatorial optimization}
\newacronym{cordic}{CORDIC}{coordinate rotation digital computer}
\newacronym{cost}{COST}{commercial off-the-shelf}
\newacronym{cots}{COTS}{commercial off-the-shelf}
\newacronym{cp}{CP}{cyclic prefix}
\newacronym{cpt}{CPT}{capacitive power transfer}
\newacronym{cpu}{CPU}{central-processing unit}
\newacronym{cpw}{CPW}{coplanar waveguide}
\newacronym{cqi}{CQI}{channel quality indicator}
\newacronym{cr}{CR}{coding rate}
\newacronym{crc}{CRC}{cyclic redundancy check}
\newacronym{crlb}{CRLB}{Cram\'er-Rao lower bound}
\newacronym{crs}{CRS}{cell reference signal}
\newacronym{cs}{CS}{compressed sensing}
\newacronym{cse}{CSE}{channel state estimation}
\newacronym{csi}{CSI}{channel state information}
\newacronym{csp}{CSP}{contact service point}
\newacronym{css}{CSS}{chirp spread spectrum}
\newacronym{cu}{CU}{central unit}
\newacronym{cv}{CV}{constant voltage}
\newacronym{cw}{CW}{continuous wave}
\newacronym{d2d}{D2D}{device-to-device}
\newacronym{dab}{DAB}{Digital Audio Broadcasting}
\newacronym{dac}{DAC}{digital-to-analog converter}
\newacronym{daq}{DAQ}{data acquisition system}
\newacronym{das}{DAS}{distributed antenna systems}
\newacronym{dbpsk}{DBPSK}{Differential Binary Phase Shift Keying}
\newacronym{dc}{DC}{direct current}
\newacronym{dcc}{DCC}{dynamic cooperation clustering}
\newacronym{ddc}{DDC}{digital down conversion}
\newacronym{de}{DE}{drain efficiency}
\newacronym{dft}{DFT}{discrete Fourier transform}
\newacronym{dl}{DL}{downlink}
\newacronym{dlc}{DLC}{Distributed Laser Charging}
\newacronym{dli}{DLI}{direct link interference}
\newacronym{dlt}{DLT}{Distributed Ledger Technology}
\newacronym{dm}{DM}{diffuse multipath}
\newacronym{dma}{DMA}{Direct Memory Access}
\newacronym{dmac}{DMAC}{Direct Memory Access Controller}
\newacronym{dmc}{DMC}{diffuse multipath component}
\newacronym{dmimo}{D-MIMO}{distributed MIMO}
\newacronym{dnn}{DNN}{deep neural network}
\newacronym{doa}{DOA}{direction-of-arrival}
\newacronym{dp}{DP}{differencial privacy}
\newacronym{dpd}{DPD}{digital pre-distortion}
\newacronym{dpdk}{DPDK}{Data Plane Development Kit}
\newacronym{dram}{DRAM}{dynamic random-access memory}
\newacronym{drcs}{$\Delta$RCS}{differential-radar cross section}
\newacronym{drx}{DRX}{Discontinuous Reception Mode}
\newacronym{dsb}{DSB}{double-sideband}
\newacronym{dsl}{DSL}{digital subscriber line}
\newacronym{dsp}{DSP}{digital signal processing}
\newacronym{dss}{DSS}{Dataset Storage Standard}
\newacronym{duc}{DUC}{digital up-converter}
\newacronym{dvb}{DVB}{Digital Video Broadcasting}
\newacronym{e2e}{E2E}{end-to-end}
\newacronym{easa}{EASA}{European Union Aviation Safety Agency}
\newacronym{ebg}{EBG}{electromagnetic bandgap}
\newacronym{ec}{EC}{European Commission}
\newacronym{ecc}{ECC}{elliptic curve cryptography}
\newacronym{ecdf}{eCDF}{empirical cumulative distribution function}
\newacronym{ecdlp}{ECDLP}{elliptic curve discrete logarithm problem}
\newacronym{ecsp}{ECSP}{edge computing service point}
\newacronym{edlc}{EDLC}{electrostatic double-layer capacitors}
\newacronym{edrx}{eDRX}{Extended Discontinuous Reception Mode}
\newacronym{ee}{EE}{energy efficiency}
\newacronym{egprs}{EGPRS}{Enhanced Data Rates for GSM Evolution}
\newacronym{eh}{EH}{energy harvesting}
\newacronym{eirp}{EIRP}{equivalent isotropically radiated power}
\newacronym{em}{EM}{electromagnetic}
\newacronym{embb}{eMBB}{enhanced Mobile Broadband}
\newacronym{en}{EN}{energy neutral}
\newacronym{end}{END}{energy-neutral device}
\newacronym{enob}{ENOB}{effective number of bits}
\newacronym{eol}{EoL}{end of life}
\newacronym{ep}{EP}{energy profiler}
\newacronym{epd}{EPD}{electronic paper display}
\newacronym{epu}{EPU}{edge processing unit}
\newacronym{er}{ER}{Energy Receiver}
\newacronym{erp}{ERP}{effective radiation power}
\newacronym[plural=ESCs,firstplural=electronic speed controllers (ESCs)]{esc}{ESC}{electronic speed control}
\newacronym{esd}{ESD}{electrostatic discharge}
\newacronym{esl}{ESL}{electronic shelf label}
\newacronym{esr}{ESR}{equivalent series resistance}
\newacronym{et}{ET}{Energy Transmitter}
\newacronym{etsi}{ETSI}{European Telecommunications Standards Institute}
\newacronym{evd}{EVD}{eigenvalue decomposition}
\newacronym{evm}{EVM}{Error Vector Magnitude}
\newacronym{ewlb}{eWLB}{Embedded Wafer Level Ball Grid Array}
\newacronym{fa}{FA}{federation anchor}
\newacronym{fair}{FAIR}{Findability, Accessibility, Interoperability, and Reuse of digital assets}
\newacronym{fcf}{FCF}{frequency correlation function}
\newacronym{fd}{FD}{front-haul distance}
\newacronym{fdd}{FDD}{frequency-division duplexing}
\newacronym{fde}{FDE}{frequency domain equalizer}
\newacronym{fdm}{FDM}{frequency-division multiplexing}
\newacronym{fdma}{FDMA}{frequency division multiple access}
\newacronym{fem}{FEM}{finite element analysis}
\newacronym{fembb}{feMBB}{further enhanced mobile broadband}
\newacronym{fft}{FFT}{fast Fourier transform}
\newacronym{fh}{FH}{fronthaul}
\newacronym{fhss}{FHSS}{frequency hopping spread spectrum}
\newacronym{fifo}{FIFO}{First In, First Out}
\newacronym{fim}{FIM}{Fisher information matrix}
\newacronym{fits}{FITS}{Flexible Image Transport System}
\newacronym{fl}{FL}{federated learning}
\newacronym{fom}{FoM}{figure of merit}
\newacronym{fov}{FOV}{field of view}
\newacronym{fpga}{FPGA}{field-programmable gate array}
\newacronym{fram}{FRAM}{Ferroelectric Random Access Memory}
\newacronym{fsk}{FSK}{frequency shift keying}
\newacronym{fspl}{FSPL}{free space path loss}
\newacronym{fss}{FSS}{frequency selective surface}
\newacronym{gb}{GB}{grant-based}
\newacronym{gdpr}{GDPR}{general data protection regulation}
\newacronym{gf}{GF}{grant-free}
\newacronym{gmsk}{GMSK}{Gaussian minimum-shift keying}
\newacronym{gnb}{gNB}{Next Generation Node B}
\newacronym{gni}{GNI}{gross national income}
\newacronym{gnn}{GNN}{graph neural network}
\newacronym{gnss}{GNSS}{global navigation satellite system}
\newacronym{gpclk}{GPCLK}{general purpose clock}
\newacronym{gpio}{GPIO}{General-Purpose Input/Output}
\newacronym{gpl}{GPL}{GNU General Public License}
\newacronym{gprs}{GPRS}{General Packet Radio Services}
\newacronym{gps}{GPS}{Global Positioning System}
\newacronym{gpu}{GPU}{graphical processing unit}
\newacronym{grc}{GRC}{GNU Radio Companion}
\newacronym{gscm}{GSCM}{geometry‐based stochastic model}
\newacronym{gsm}{GSM}{Global System for Mobile Communications}  %
\newacronym{gwp}{GWP}{Global Warming Potential}
\newacronym{harq}{HARQ}{hybrid automatic repeat request}
\newacronym{hat}{HAT}{hardware attached on top}
\newacronym{hcs}{HCS}{human-centric services}
\newacronym{hdf5}{HDF5}{Hierarchical Data Format version 5}
\newacronym{hfss}{HFSS}{High Frequency Simulator Software}
\newacronym{hmd}{HMD}{head-mounted display}
\newacronym{hpbm}{HPBM}{half power beam width}
\newacronym{HyMPRo}{HyMPRo}{Hybrid Multi-Path Routing algorithm}
\newacronym{i.i.d.}{i.i.d.}{independent and identically distributed}
\newacronym{i2c}{I2C}{Inter-Integrated Circuit}
\newacronym{iaq}{IAQ}{Indoor Air Quality}
\newacronym{ib}{IB}{in-band}
\newacronym{ibbc}{IBBC}{inter-band beam configuration}
\newacronym{ibo}{IBO}{input back-off}
\newacronym{ic}{IC}{integrated circuit}
\newacronym{iccs}{ICCS}{Ilmsens correlative channel sounder}
\newacronym{ici}{ICI}{intercarrier interference}
\newacronym{icnirp}{ICNIRP}{International Commission on Non-Ionizing Radiation Protection}
\newacronym{id}{ID}{information decoding}
\newacronym{idft}{IDFT}{inverse discrete Fourier transform}
\newacronym{if}{IF}{intermediate-frequency}
\newacronym{iid}{i.i.d.}{independently and identically distributed}
\newacronym{iis}{IIS}{integrated information system}
\newacronym{im}{IM}{intermodulation}
\newacronym{imd}{IMD}{intermodulation distortion}
\newacronym{imu}{IMU}{inertial measurement unit}
\newacronym{inh}{InH}{indoor hotspot office}
\newacronym{io}{IO}{input/output}
\newacronym{ioe}{IoE}{Internet of Everything}
\newacronym{iot}{IoT}{Internet of Things}
\newacronym{ipt}{IPT}{inductive power transfer}
\newacronym{ipy}{IPY}{Interventions per Year}
\newacronym{iq}{IQ}{in-phase and quadrature}
\newacronym{iqi}{IQI}{IQ imbalance}
\newacronym{ir}{IR}{infrared}
\newacronym{isi}{ISI}{intersymbol interference}
\newacronym{ism}{ISM}{industrial, scientific and medical}
\newacronym{isp}{ISP}{internet service provider}
\newacronym{itu}{ITU}{International Telecommunication Union}
\newacronym{jesd}{JESD}{Joint Electron Devices Engineering Council}
\newacronym{jfet}{JFET}{junction field effect transistor}
\newacronym{kpi}{KPI}{key performance indicator}
\newacronym{kvi}{KVI}{key value indicator}
\newacronym{larva}{LARVA}{LARge Virtual Array}
\newacronym{lca}{LCA}{life cycle assessment}
\newacronym{lco}{LCO}{lithium cobalt oxide}
\newacronym{ldo}{LDO}{Low-dropout voltage regulator}
\newacronym{ldpc}{LDPC}{low-density parity-check}
\newacronym{led}{LED}{Light Emitting Diode}
\newacronym{less}{LESS}{Low Energy Scheduler Solution}
\newacronym{lfp}{LFP}{lithium iron phosphate}
\newacronym{lib}{LIB}{Lithium-Ion Battery}
\newacronym{lic}{LIC}{lithium-ion capacitor}
\newacronym{lid}{LID}{Lithium Iron Disulfide}
\newacronym{lidar}{LiDAR}{light detection and ranging}
\newacronym{liion}{Li-ion}{lithium-ion}
\newacronym{lipo}{LiPo}{lithium polymer}
\newacronym{lis}{LIS}{large intelligent surface}
\newacronym{llh}{LLH}{log-likelihood}
\newacronym{lmd}{LMD}{Lithium Manganese Dioxide}
\newacronym{lmmse}{LMMSE}{least minimum mean square error}
\newacronym{lmo}{LMO}{lithium ion manganese oxide}
\newacronym{lna}{LNA}{low-noise amplifier}
\newacronym{lo}{LO}{local oscillator}
\newacronym{lora}{LoRa}{long range}
\newacronym{lorawan}{LoRaWAN}{long-range wide-area network}
\newacronym{los}{LoS}{line-of-sight}
\newacronym{lp}{LP}{linear programming}
\newacronym{lpf}{LPF}{low-pass filter}
\newacronym{lpt}{LPT}{laser power transfer}
\newacronym{lpwa}{LPWA}{Low Power Wide Area}
\newacronym{lpwan}{LPWAN}{low-power wide-area network}
\newacronym{lpwans}{LPWANs}{Low-Power Wide-Area Networks}
\newacronym{lqi}{LQI}{link quality indicator}
\newacronym{lrelu}{LReLU}{leaky rectified linear unit}
\newacronym{lrt}{LRT}{likelihood-ratio test}
\newacronym{ls}{LS}{least squares}
\newacronym{lsa}{LSA}{large synthetic array}
\newacronym{lsf}{LSF}{large-scale fading}
\newacronym{lsfc}{LSFC}{large-scale fading component}
\newacronym{lstm}{LSTM}{Long Short-Term Memory}
\newacronym{ltc}{LTC}{lithium thionyl chloride}
\newacronym{lte}{LTE}{Long Term Evolution}
\newacronym{lti}{LTI}{linear time-invariant}
\newacronym{lto}{LTO}{lithium titanate}
\newacronym{lusta}{LUSTA 5G }{Logistique mUltimodale Sécuritaire Téléopérée \& Autonome 5G}
\newacronym{m2m}{M2M}{machine to machine}
\newacronym{mac}{MAC}{Medium Access Control}
\newacronym{mate}{MATE}{millimeter-wave MIMO testbed}
\newacronym{mc}{MC}{Monte Carlo}
\newacronym{mcl}{MCL}{Maximum Coupling Loss}
\newacronym{mcs}{MCS}{modulation and coding scheme}
\newacronym{mcu}{MCU}{microcontroller unit}
\newacronym{mec}{MEC}{multi-access edge computing}
\newacronym{mems}{MEMS}{micro-electromechanical systems}
\newacronym{mf}{MF}{matched filter}
\newacronym{mimo}{MIMO}{multiple-input multiple-output}
\newacronym{miso}{MISO}{multiple-input single-output}
\newacronym{ml}{ML}{machine learning}
\newacronym{mlp}{MLP}{multilayer perceptron}
\newacronym{mmic}{MMIC}{monolithic microwave integrated circuit}
\newacronym{mmimo}{mMIMO}{massive MIMO}
\newacronym{mmse}{MMSE}{minimum mean square error}
\newacronym{mmtc}{mMTC}{massive machine-typed communication}
\newacronym{mmwave}{mmWave}{millimeter wave}
\newacronym{mn}{MN}{matching network}
\newacronym{mosfet}{MOSFET}{metal-oxide semiconductor field effect transistor}
\newacronym{mpc}{MPC}{multipath component}
\newacronym{mppt}{MPPT}{maximum power point tracking}
\newacronym{mr}{MR}{maximum ratio}
\newacronym{mrc}{MRC}{maximum ratio combining}
\newacronym{mrc_em}{MRC}{maximum ratio combining}
\newacronym{mrc_EM}{MRC}{Magnetic Resonance Coupling}
\newacronym{mrt}{MRT}{maximum ratio transmission}
\newacronym{mse}{MSE}{mean square error}
\newacronym{msk}{MSK}{Minimum-Shift Keying}
\newacronym{mtc}{MTC}{Machine-Type Communication}
\newacronym{multi-rat}{Multi-RAT}{multiple radio access technology}
\newacronym{multirat}{Multi-RAT}{Multiple Radio Access Technology}
\newacronym{music}{MUSIC}{MUltiple SIgnal Classification}
\newacronym{navauwall}{NAVAUWALL}{AUtomated NAVigation in WALLonia}
\newacronym{nb}{NB}{narrowband}
\newacronym{nbiot}{NB-IoT}{narrowband IoT}
\newacronym{nca}{NCA}{nickel cobalt aluminum}
\newacronym{netcdf}{NetCDF}{Network Common Data Form}
\newacronym{nf}{NF}{noise figure}
\newacronym{nfc}{NFC}{near-field communication}
\newacronym{nfv}{NFV}{network function virtualization}
\newacronym{ngmn}{NGMN}{Next Generation Mobile Networks }
\newacronym{ni}{NI}{National Instruments}
\newacronym{nicd}{NiCd}{nikkel cadmium}
\newacronym{nimh}{NiMH}{nikkel metal hydride}
\newacronym{nlos}{NLoS}{non-line-of-sight}
\newacronym{nmc}{NMC}{nickel manganese cobalt}
\newacronym{nmos}{nMOS}{n-channel metal-oxide semiconductor}
\newacronym{nn}{NN}{neural network}
\newacronym{nnls}{NNLS}{non-negative least squares}
\newacronym{noma}{NOMA}{non-orthogonal multiple access}
\newacronym{np}{NP}{Neyman-Pearson}
\newacronym{npbch}{NPBCH}{Narrowband Physical Broadcast Channel}
\newacronym{npss}{NPSS}{Narrow Band Primay Synchronization Signal}
\newacronym{nr}{NR}{New Radio}
\newacronym{nrs}{NRS}{Narrow Band Reference Signal}
\newacronym{nsss}{NSSS}{Narrowband Secondary Synchronization Signal}
\newacronym{ntp}{NTP}{network time protocol}
\newacronym{oai}{OAI}{OpenAirInterface} % no spelling mistake, it is spelled all glued to eachother...
\newacronym{obw}{OBW}{occupied bandwidth}
\newacronym{ofdm}{OFDM}{orthogonal frequency-division multiplexing}
\newacronym{ofdma}{OFDMA}{orthogonal frequency-division multiple access}
\newacronym{ofdmim}{OFDM-IM}{OFDM with index modulation}
\newacronym{olos}{OLoS}{obstructed-line-of-sight}
\newacronym{oma}{OMA}{orthogonal multiple access}
\newacronym{oob}{OOB}{out-of-band}
\newacronym{ook}{OOK}{on-off keying}
\newacronym{oran}{O-RAN}{open radio-access network}
\newacronym{os}{OS}{operating system}
\newacronym{ota}{OTA}{over-the-air}
\newacronym{otaa}{OTAA}{over-the-air authentication}
\newacronym{p1}{P1}{Phase 1}
\newacronym{p2}{P2}{Phase 2}
\newacronym{p2p}{P2P}{point-to-point}
\newacronym{pa}{PA}{power amplifier}
\newacronym{pae}{PAE}{power-added efficiency}
\newacronym{pana}{PanA}{Panel A}
\newacronym{panb}{PanB}{Panel B}
\newacronym{papr}{PAPR}{peak-to-average power ratio}
\newacronym{pc}{PC}{pilot count}
\newacronym{pcb}{PCB}{printed circuit board}
\newacronym{pcg}{PCG}{power consumption gain}
\newacronym{pcie}{PCIe}{Peripheral Component Interconnect Express}
\newacronym{pcsi}{PCSI}{perfect channel state information}
\newacronym{pd}{PD}{powered device}
\newacronym{pdcch}{PDCCH}{physical downlink control channel}
\newacronym{pdf}{PDF}{probability density function}
\newacronym{pdp}{PDP}{power delay profile}
\newacronym{pdsch}{PDSCH}{physical downlink shared channel}
\newacronym{pe}{PE}{processing element}
\newacronym{peb}{PEB}{positioning error bound}
\newacronym{per}{PER}{packet error rate}
\newacronym{pet}{PET}{privacy enhancing technology}
\newacronym{pg}{PG}{path gain}
\newacronym{pgd}{PGD}{proximal gradient descent}
\newacronym{phy}{PHY}{physical}
\newacronym{pki}{PKI}{public key infrastucture}
\newacronym{pl}{PL}{path loss}
\newacronym{pla}{PLA}{physically large array}
\newacronym{pll}{PLL}{phase-locked loop}
\newacronym[plural=PMs,firstplural=person months (PMs)]{pm}{PM}{person month}
\newacronym{pmf}{PMF}{polymer microwave fiber}
\newacronym{pmu}{PMU}{power management unit}
\newacronym{pn}{PN}{pseudo-noise}
\newacronym{poc}{PoC}{proof of concept}
\newacronym{poe}{PoE}{power-over-Ethernet}
\newacronym{pps}{1PPS}{pulse per second}
\newacronym{pr}{PR}{phase reversal}
\newacronym{prb}{PRB}{Physical Resource Block}
\newacronym{prbs}{PRBs}{Physical Resource Blocks}
\newacronym{prs}{PRS}{Peripheral Reflex System}
\newacronym{ps}{PS}{Processing System}
\newacronym{psd}{PSD}{power spectral density}
\newacronym{pse}{PSE}{power sourcing equipment}
\newacronym{psk}{PSK}{phase shift keying}
\newacronym{psm}{PSM}{power saving mode}
\newacronym{pss}{PSS}{primary synchronisation signal}
\newacronym{ptp}{PTP}{precision-time protocol}
\newacronym{ptrs}{PTRS}{Phase-Tracking Reference Signals}
\newacronym{ptw}{PTW}{paging time window}
\newacronym{pv}{PV}{photovoltaic}
\newacronym{pw}{PW}{planar wavefront}
\newacronym{pwm}{PWM}{pulse width modulation}
\newacronym{qam}{QAM}{quadrature amplitude modulation}
\newacronym{qos}{QoS}{quality-of-service}
\newacronym{qpsk}{QPSK}{quadrature phase-shift keying}
\newacronym{qrrls}{QR-RLS}{QR decomposition based recursive least squares}
\newacronym{quadriga}{QuaDRiGa}{QUAsi Deterministic RadIo channel GenerAtor}
\newacronym{ra}{RA}{Random Access}
\newacronym{ram}{RAM}{random-access memory}
\newacronym{ran}{RAN}{radio access network}
\newacronym{rar}{RAR}{Random Access Response}
\newacronym{rat}{RAT}{radio access technology}
\newacronym{rb}{Rb}{Rubidium}
\newacronym{rbs}{RBS}{radio base station}
\newacronym{rbw}{RBW}{resolution bandwidth}
\newacronym{rcs}{RCS}{radar cross section}
\newacronym{rdl}{RDL}{redistribution layer}
\newacronym{re}{RE}{radio element}
\newacronym{relu}{ReLU}{rectified linear unit}
\newacronym{rf}{RF}{radio frequency}
\newacronym{rfeh}{RFEH}{radio frequency energy harvesting}
\newacronym{rfic}{RFIC}{radio-frequency integrated circuit}
\newacronym{rfid}{RFID}{radio frequency identification}
\newacronym{rfpt}{RFPT}{radio frequency power transfer}
\newacronym{rfsoc}{RFSoC}{Radio Frequency System-on-Chip}
\newacronym{rir}{RIR}{room impulse response}
\newacronym{ris}{RIS}{reflective intelligent surface}%reconfigurable?
\newacronym{rllmtc}{RLLMTC}{reliable low latency machine type communication}
\newacronym{rls}{RLS}{recursive least squares}
\newacronym{rms}{RMS}{root-mean-square}
\newacronym{rmse}{RMSE}{root-mean-square error}
\newacronym{rmt}{RMT}{random matrix theory}
\newacronym{rof}{RoF}{radio-over-fiber}
\newacronym{ros}{ROS}{robot operating system}
\newacronym{rpi}{RPi}{Raspberry Pi}
\newacronym{rrc}{RRC}{Radio Resource Connection}
\newacronym{rreq}{RREQ}{route request packet}
\newacronym{rsrp}{RSRP}{Reference Signals Received Power}
\newacronym{rsrq}{RSRQ}{Reference Signal Received Quality}
\newacronym{rss}{RSS}{received signal strength}
\newacronym{rssi}{RSSI}{received signal strength indicator}
\newacronym{rtc}{RTC}{real time clock}
\newacronym{rtf}{RTF}{reader talks first}
\newacronym{rtk}{RTK}{real time kinematics}
\newacronym{rts}{RTS}{ray tracing simulator}
\newacronym{ru}{RU}{Radio Unit}
\newacronym{rv}{RV}{random variable}
\newacronym{rw}{RW}{RadioWeaves}
\newacronym{rx}{RX}{receiver}
\newacronym{rzf}{RZF}{regularized zero forcing}
\newacronym{s-parameter}{S-parameter}{scattering parameter}
\newacronym{sa}{SA}{synchronization anchor}
\newacronym{sa5}{SA}{Stand Alone}
\newacronym{sar}{SAR}{specific absorption rate}
\newacronym{sbl}{SBL}{sparse Bayesian learning}
\newacronym{scfdma}{SCFDMA}{single-carrier frequency division multiple access}
\newacronym{sdg}{SDG}{Sustainable Development Goal}
\newacronym{sdm}{SDM}{sigma-delta modulator}
\newacronym{sdma}{SDMA}{spatial-division multiple access}
\newacronym{sdn}{SDN}{software-defined network}
\newacronym{sdof}{SDoF}{sigma-delta over fiber}
\newacronym{sdr}{SDR}{software-defined radio}
\newacronym{se}{SE}{spectral efficiency}
\newacronym{sei}{SEI}{specific emitter identification}
\newacronym{ser}{SER}{symbol-error rate}
\newacronym{sf}{SF}{spreading factor}
\newacronym{sfn}{SFN}{single frequency network}
\newacronym{sfp}{SFP}{small form-factor pluggable}
\newacronym{sha}{SHA}{Secure Hash Algorithm}
\newacronym{SigMF}{SigMF}{Signal Metadata Format}
\newacronym{simo}{SIMO}{single-input multiple-output}
\newacronym{sinr}{SINR}{signal-to-interference-plus-noise ratio}
\newacronym{siso}{SISO}{single-input single-output}
\newacronym{slam}{SLAM}{simultaneous localization and mapping}
\newacronym{slc}{SLC}{spatial leakage suppression}
\newacronym{slerp}{SLERP}{spherical linear interpolation}
\newacronym{sma}{SMA}{SubMiniature version A}
\newacronym{smc}{SMC}{specular multipath component}
\newacronym{smps}{SMPS}{switched mode power supply}
\newacronym{sndr}{SNDR}{signal-to-noise-and-distortion ratio}
\newacronym{snidr}{SNIDR}{signal-to-noise-and-interference-and-distortion ratio}
\newacronym{snir}{SNIR}{signal-to-interference-plus-noise ratio}
\newacronym{snr}{SNR}{signal-to-noise ratio}
\newacronym{soc}{SoC}{state of charge}
\newacronym{SoC}{SOC}{System on Chip}
\newacronym{sp}{SP}{service point}
\newacronym{spdt}{SPDT}{single pole double throw}
\newacronym{spi}{SPI}{Serial Peripheral Interface}
\newacronym{spst}{SPST}{single pole single throw}
\newacronym{sram}{SRAM}{static random-access memory}
\newacronym{srd}{SRD}{short-range device}
\newacronym{srls}{SRLS}{standard recursive least squares}
\newacronym{srs}{SRS}{Sounding Reference Signal}
\newacronym{ssb}{SSB}{synchronisation signal block}
\newacronym{ssd}{SSD}{solid state drive}
\newacronym{ssq}{SSQ}{simulator sickness questionnaire}
\newacronym{steam}{STEAM}{science, technology, engineering, the arts, and mathematics}
\newacronym{svd}{SVD}{singular value decomposition}
\newacronym{sw}{SW}{spherical wavefront}
\newacronym{swipt}{SWIPT}{simultaneous wireless information and power transfer}
\newacronym{synce}{SyncE}{Synchronous Ethernet}
\newacronym{ta}{TA}{timing advance}
\newacronym{tau}{TAU}{tracking area update}
\newacronym{tcer}{TCER}{transported to consumed energy ratio}
\newacronym{tcp}{TCP}{Transmission Control Protocol}
\newacronym{tcxo}{TCXO}{temperature compensated crystal oscillator}
\newacronym{tdd}{TDD}{time division duplexing}
\newacronym{tdma}{TDMA}{time division-multiple access}
\newacronym{tdoa}{TDOA}{time-difference-of-arrival}
\newacronym{thz}{THz}{Terahertz}
\newacronym{to}{TO}{timing offset}
\newacronym{toa}{TOA}{time-of-arrival}
\newacronym{tof}{ToF}{time-of-flight}
\newacronym{tosm}{TOSM}{through-open-short-match}
\newacronym{tpms}{TPMS}{Tire-Pressure Monitoring System}
\newacronym{trl}{TRL}{technology readyness level}
\newacronym{trp}{TRP}{Transmission Reception Point}
\newacronym{tsn}{TSN}{time-sensitive networking}
\newacronym{ttf}{TTF}{tag talks first}
\newacronym{ttff}{TTFF}{Time To First Fix}
\newacronym{ttm}{TTM}{time to market}
\newacronym{ttn}{TTN}{The Things Network}
\newacronym{tx}{TX}{transmitter}
\newacronym{uart}{UART}{Universal Asynchronous Receiver/Transmitter}
\newacronym{uav}{UAV}{unmanned aerial vehicle}
\newacronym{uc}{UC}{use case}
\newacronym{ucie}{UCIe}{Universal Chiplet Interconnect Express}
\newacronym{udp}{UDP}{User Datagram Protocol}
\newacronym{ue}{UE}{user equipment}
\newacronym{ugv}{UGV}{unmanned ground vehicle}
\newacronym{uhd}{UHD}{USRP hardware driver}
\newacronym{uhf}{UHF}{ultra-high frequency}
\newacronym{ul}{UL}{uplink}
\newacronym{ula}{ULA}{uniform linear array}
\newacronym{uMIMO}{$\mu$-MIMO}{ultra-massive Multiple-Input Multiple-Output}
\newacronym{ummtc}{umMTC}{ultra massive machine type communication}
\newacronym{un}{UN}{United Nations}
\newacronym{upa}{UPA}{uniform planar array}
\newacronym{ura}{URA}{uniform rectangular array}
\newacronym{urllc}{URLLC}{ultra-reliable low-latency communications}
\newacronym{usrp}{USRP}{universal software radio peripheral}
\newacronym{uv}{UV}{unmanned vehicle}
\newacronym{uwb}{UWB}{ultrawideband}
\newacronym{v2v}{V2V}{vehicle-to-vehicle}
\newacronym{vco}{VCO}{voltage-controlled oscillator}
\newacronym{vep}{VEP}{virtual edge platform}
\newacronym{vlc}{VLC}{visible light communication}
\newacronym{vlp}{VLP}{visible light positioning}
\newacronym{vna}{VNA}{vector network analyzer}
\newacronym{voc}{VOC}{Voltatile Organic Compound}
\newacronym{votable}{VOTable}{Virtual Observatory Table}
\newacronym{vr}{VR}{virtual reality}
\newacronym{vuca}{VUCA}{volatile, uncertain, complex and ambiguous}
\newacronym{wb}{WB}{wideband}
\newacronym{wban}{WBAN}{wireless body area network}
\newacronym{wd}{WD}{Wireless distance}
\newacronym{wimax}{WiMAX}{Worldwide Interoperability for Microwave Access}
\newacronym{wlan}{WLAN}{wireless LAN}
\newacronym[plural=WPs,firstplural=work packages (WPs)]{wp}{WP}{work package}
\newacronym{wpt}{WPT}{wireless power transfer}
\newacronym{wr}{WR}{White Rabbit}
\newacronym{wrsn}{WRSN}{wireless rechargeable sensor network}
\newacronym{wsn}{WSN}{wireless sensor network}
\newacronym{xets}{XETS}{cross exponentially tapered slot}
\newacronym{xlmimo}{XL-MIMO}{extremely large-scale MIMO}
\newacronym{xr}{XR}{extended reality}
\newacronym{z3ro}{Z3RO}{zero third-order distortion}
\newacronym{zf}{ZF}{zero-forcing}
\newacronym{zmcscg}{ZMCSCG}{zero mean circularly symmetric complex Gaussian}
\newacronym{zmq}{ZMQ}{ZeroMQ}

\newacronym{ghg}{GHG}{greenhouse gas}
\newacronym{llm}{LLM}{Large Language Model}

\newacronym{ric}{RIC}{Radio Intelligent Controller}

\newacronym{smp}{SMP}{Sustainability Management Plane}
\newacronym{imt}{IMT}{International Mobile Telecommunication}

\newacronym{aon}{AON}{All-Optical Network}
\newacronym{emf}{EMF}{electromagnetic field}
\newacronym{ha}{HA}{Hardware Accelerator}
\newacronym{sdo}{SDO}{standards developing organization}
\usepackage{siunitx}
\usepackage{tikz, pgfplots}
\usetikzlibrary{backgrounds}
\usetikzlibrary{calc}
\usepackage[capitalize]{cleveref}
\usepackage[most]{tcolorbox}
\usepackage{lipsum}  
\usepackage{mdframed}
\usepackage{tabularx} 
\usepackage{booktabs}
\usepackage{tikz}
\newcommand{\circnum}[1]{%
  \tikz[baseline=(char.base)]{
    \node[shape=circle,draw,inner sep=1pt] (char) {#1};}}
    
\setlipsum{%
  par-before = \begingroup\color{gray},
  par-after = \endgroup
}

\def\BibTeX{{\rm B\kern-.05em{\sc i\kern-.025em b}\kern-.08em
    T\kern-.1667em\lower.7ex\hbox{E}\kern-.125emX}}

\newcommand{\update}[1]{\textcolor{black}{#1}}

\begin{document}

\title{Towards Sustainable 6G: A Holistic View of Trade-offs and Enablers}

\author{Mattia Merluzzi, Olivier Bouchet, Ali Balador, Gilles Callebaut, Anastasius Gavras, \\ Liesbet Van der Perre, Albert Banchs, Mauro Renato Boldi, Emilio Calvanese Strinati,\\ Bahare M Khorsandi, Marja Matinmikko-Blue, Lars Christoph Schmelz
        % <-this % stops a space
\thanks{Mattia Merluzzi and Emilio Calvanese Strinati are with Univ. Grenoble Alpes, CEA, Leti, France; Olivier Bouchet is with Orange Labs, France; Ali Balador is with Ericsson Research, Sweden; Gilles Callebaut and Liesbet Van der Perre are with KU Leven, Belgium; Anastasius Gavras is with Eurescom, Germany; Albert Banchs is with University Carlos III and IMDEA Networks institute, Spain; Mauro Renato Boldi is with Telecom Italia, Italy; Bahare M. Khorsandi and Lars Christoph Schmelz are with Nokia Germany; Marja Matinmikko-Blue is with University of Oulu, Finland. \\
This work is co-funded by the European Union under Grant Agreement 101191936. Views and opinions expressed are however those of the authors only and do not necessarily reflect those of all SUSTAIN-6G consortium parties nor those of the European Union or the SNS JU (granting authority). Neither the European Union nor the granting authority can be held responsible for them. The authors would like to thank the whole SUSTAIN-6G consortium. Figs.~\ref{fig:tradeoffs}, \ref{fig:6G_KVIs} were partly designed with Flaticon.com resources.
}% <-this % stops a space
%\thanks{Manuscript received April 19, 2021; revised August 16, 2021.}
}



\maketitle

\begin{abstract}
The sixth generation of mobile networks (6G) can play a central role in shaping a sustainable future, the most compelling contemporary challenge. Connecting the unconnected, reducing carbon emissions of vertical sectors, and allowing heterogeneous types of intelligence (including humans) to safely and constructively interact in complex environments, are only a few of the several challenges that can be supported by 6G. However, this requires a careful design that balances positive and negative impacts of 6G, towards a sustainable and sustainability-enabling technology. This paper presents a holistic view that translates the complex interplay between the 6G enabling effects and the sustainability of 6G by design, into concrete trade-offs and research questions. Starting from today's challenges for society and associated key values, we unfold the dilemma into a set of technical trade-offs, whose solutions span from technological innovations to standardization actions towards applicability.
\end{abstract}

\begin{IEEEkeywords}
Sustainability, 6G, holistic view, eco-innovation, Sustainable 6G, 6G for Sustainability
\end{IEEEkeywords}

\section{Introduction}\label{sec:intro}

Today, in line with the United Nations' \glspl{sdg}~\cite{UN701}, achieving sustainability is a fundamental objective to address a broad spectrum of contemporary challenges \cite{EAZ25}, ranging from social deficiencies and the environmental crisis to the imperatives of economic development. Sustainability means ``meeting the needs of the present without compromising the ability of future generations to meet their own needs.''\cite{UNsustainability}
The actions towards this global objective orbit around three \textit{sustainability pillars}: \textbf{\textit{environmental, social, and economic}}, respectively related to planet, people, and profit. \update{These intricately interrelated pillars require a \textit{holistic} view, resulting into a complex \textbf{\textit{overarching dilemma}} between positive and negative impacts of every action.}  

Wireless technology innovation is not dispensed from these challenges. The development of the \gls{6g} of mobile networks represents a perfect example of this dilemma. On one hand, \gls{6g} comes with potentially negative impacts including, among others, material use and energy consumption (environmental), potentially harmful impacts of technology on people (social), capital and operational expenditure for operators (economic). On the other hand, \gls{6g} comes with potentially positive impacts, enabling other sectors to: \textit{i)} reduce their energy consumption, \gls{ghg} emissions, and resource usage (environmental); \textit{ii)} bridge the digital divide by enabling inclusive and trustworthy access to technologies and applications (social); \textit{iii)} enable digital transformation of businesses or create opportunities for new business models (economic). Minimizing the negative impact of 6G and maximizing its positive impact can be summarized into two respective aspects: \textbf{\textit{Sustainable 6G} (S6G)}  and \textbf{\textit{\gls{6g} for Sustainable Applications} (6GS)}. 

To analyze and address the three pillars across the two aspects (which results in \textbf{six sustainability dimensions}), all relevant stakeholders must participate in the development of \gls{6g}. 
Optimizing single segments independently (\textit{isolation of domains}), and without a view on the benefits and drawbacks for the enabled usage scenarios is not an option to achieve a \textbf{sustainable 6G by design}. Finally, 
when jointly considering the three pillars, and especially social sustainability, questions on \textit{values} beyond performance and \glspl{kpi} arise. This requires new models to measure such values, introducing \glspl{kvi}.
\update{The aim of this work is to provide a view of concrete directions to address this complex problem, from the overarching dilemma to technical solutions. Our contributions are summarized as follows:
\begin{itemize}
    \item[\textbf{(1)}] \textbf{Overarching dilemma and associated terminology:} we analyze the dilemma between positive and negative impacts of 6G, proposing a comprehensive terminology that includes all sustainability pillars, beyond the environmental one, typical of the green networking paradigm;
    \item[\textbf{(2)}] \textbf{Value-driven 6G and emerging use cases:} we discuss the framework of \textit{beyond performance towards values-driven} design, and present a set of usage scenarios (or, verticals) that are representative of the ecosystem of actors exploiting the power of \gls{ict} to generate environmental, social and economic values;
    \item[\textbf{(3)}] \textbf{From the dilemma to solutions:} advocating a holistic perspective across sustainability pillars, we break down the dilemma into concrete technical trade-offs, and propose technological innovations to explore these trade-offs;
    \item[\textbf{(4)}] \textbf{From optimality to applicability:} we present our view towards the inclusion of sustainability in \gls{6g} standardization, with examples and current status in different bodies.
\end{itemize}
The remainder of this work is organized as follows: in \cref{sec:dilemma}, we described the overarching dilemma; in \cref{sec:all_tradeoffs}, we introduce the key technical trade-offs that should be explored towards solving the dilemma; tackling these trade-offs requires a set of technology enablers, introduced in \cref{sec:tech_enablers}; further, in \cref{sec:standardization} we discuss relevant ongoing activities to make sustainability as part of 6G standardization; finally \cref{sec:conclusions} concludes the work with some takeaways.
}

\section{The overarching dilemma}\label{sec:dilemma}
\begin{figure}[t!]
    \centering
    \includegraphics[width=0.9\columnwidth]{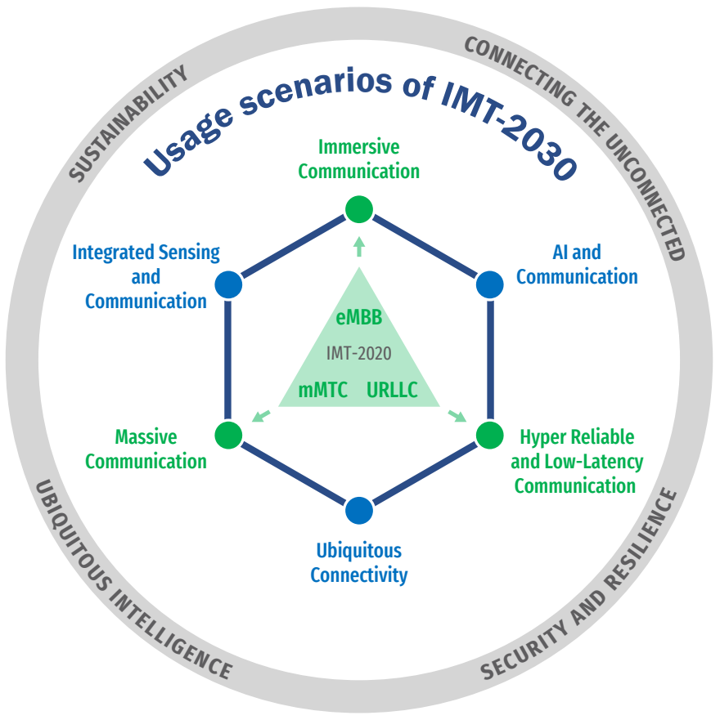}
    \caption{The ITU's IMT-2030 framework~\cite{IMT2030}, showing potentially contradicting goals. Use scenarios will further diversify.} 
    \label{fig:6Gwheel}
    %\vspace{-0.3 cm}
\end{figure}

Mobile network deployment has logically aimed economic benefits over others. 
The \gls{itu} has approved the `Framework and overall objectives of the future development of \gls{imt} for 2030 and beyond'~\cite{IMT2030}, which lays out the technical goals and key values for \gls{6g}. It clarifies the increased diversity in usage scenarios in combination with better performance by stretching and expanding the `5G triangle' to a `6G hexagon', as shown in \cref{fig:6Gwheel}. It furthermore introduces four `overarching aspects' that are sketched as an outer circle embracing the more technology-centric terminology in the core of the diagram. These aspects, which act as design principles, can be interpreted as \textbf{values} to be considered in conjunction with the performance improvements, and are:
\begin{itemize}
    \item \textbf{Sustainability}: to be understood as much broader than mere energy efficiency or even ecological footprint;
    \item \textbf{Connecting the unconnected}: to include affordable connectivity to all, including sparsely populated areas;
    \item \textbf{Ubiquitous intelligence}: \update{to profit from the exploding possibilities of computing capabilities and AI;}
    \item \textbf{Security/privacy/resilience}: to highlight the risks of relying on wireless networks in personal and public contexts;
\end{itemize}
We notice that these overarching aspects present potentially contradicting goals. e.g., resilience might call for additional infrastructure deployment, thus potentially generating negative impacts in terms of economic and environmental sustainability. These contradictions complicate the evaluation of positive and negative impacts generating from the introduction of new technologies, including 6G. 
In the context of \gls{6g} and the multi-dimensional problem space of sustainability, this paper advocates the use of the terms \textbf{\textit{first order effect}} and \textbf{\textit{second order effect}}, to describe the direct impact from the implementation of ICT and the indirect impact from its use, respectively. Both can be positive and negative. These concepts are expanded beyond the current ITU-T definition~\cite{ITU}, to allow for the joint incorporation of not exclusively environmental, but also social and economic sustainability pillars. \update{The main challenge is to solve the dilemma of \textbf{6G as part of the problem} (i.e., the need for S6G) and \textbf{6G as part of the solution} (i.e., 6GS), which is reflected in the balance between (negative) first and (positive) second order effects. This can be summarized in the following type of question for 6G:}
\begin{tcolorbox}[colframe=blue!10, colback=blue!10, coltitle=red, coltext=black]
\update{
What is the acceptable price to pay (considering environmental, social and economic sustainability pillars) in terms of negative effects, to enable positive effects?}
\end{tcolorbox}

\subsection{The $3\times2=$ 6(G) dimensions of sustainability}
\begin{figure*}[t]
    \centering
\includegraphics[width=0.98\textwidth]{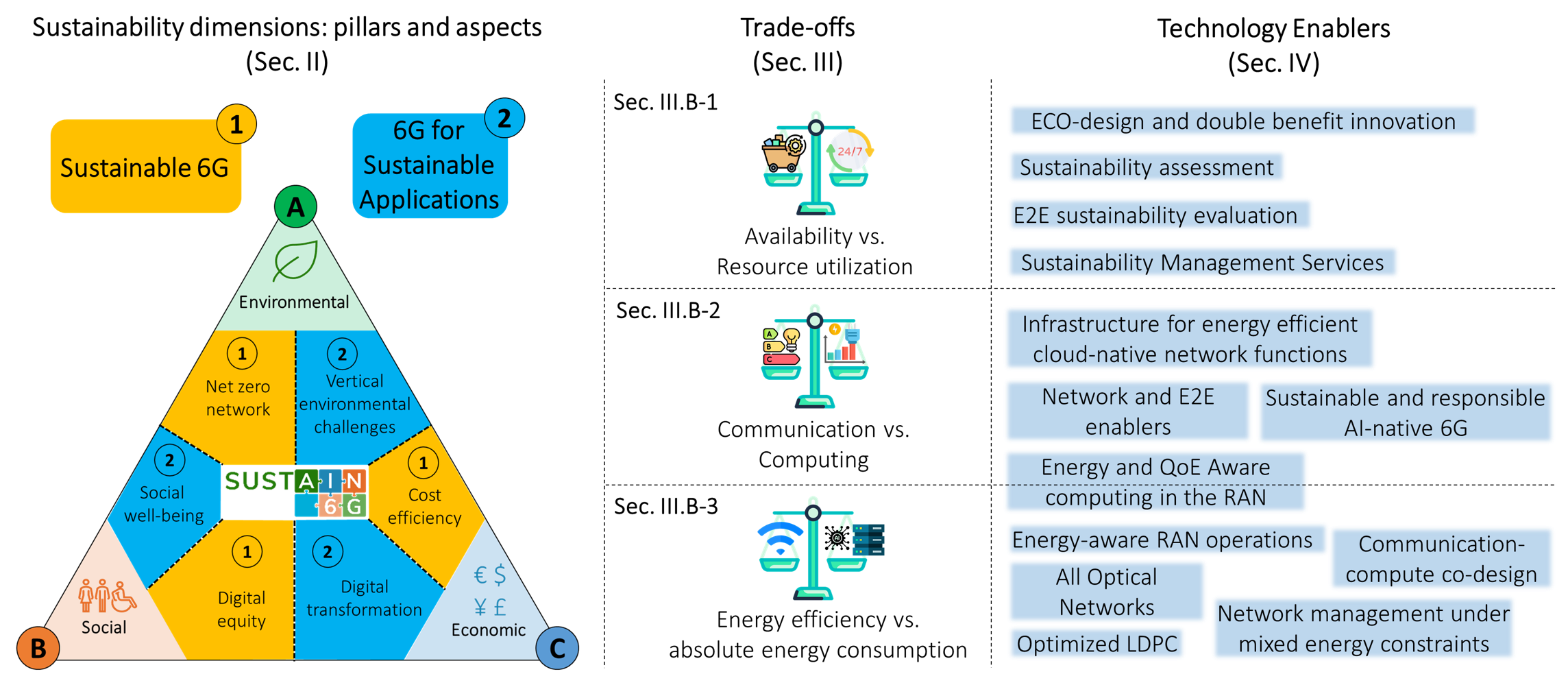}
    \caption{The six sustainability dimensions (triangle), building on a set of trade-offs (introduced in \cref{sec:all_tradeoffs}) and technology enablers (introduced in \cref{sec:tech_enablers}) of the SUSTAIN-6G project (https://sustain-6g.eu/)}
    \label{fig:tradeoffs}
    %\vspace{-0.3 cm}
\end{figure*}

Developing a 6G sustainability vision necessitates addressing the goals set within all three pillars of sustainability. Also, two critical \textbf{\textit{sustainability aspects}} must be addressed: S6G and 6GS. 
It becomes evident that a truly sustainable 6G ecosystem must comprehensively address all three pillars, across both aspects. This integrated approach results in six \textit{sustainability dimensions}.
Further, the diverse sustainability needs originating from \update{6G network operators from one hand and vertical sectors on the other hand, cannot be addressed in isolation, due to potentially conflicting objectives.} %a wide range 
For instance, expanding coverage in rural areas could support telemedicine or agriculture applications, leading to improved healthcare, reduced car travel, and decreased use of fertilizers and water. However, this leads to potentially higher energy, resource consumption \update{and monetary costs} by the communication system. \update{ \cref{fig:tradeoffs} summarizes the multi-dimensional problem of 6G and sustainability. The left hand side illustrates the six introduced sustainability dimensions, with examples of S6G and 6GS. The vertices of the triangle represent the sustainability pillars (environmental, social, and economic), respectively indicated by letters \circnum{A}, \circnum{B} and \circnum{C}.  The two sustainability aspects (S6G and 6GS) are indicated on top by numbers \circnum{1} and \circnum{2}, respectively. Also, they appear at the center of the triangle (along with their respective numbers) with examples related to the three sustainability pillars. Going further, at the center of \cref{fig:tradeoffs} we illustrate key trade-offs (introduced in \cref{sec:all_tradeoffs} to technically tackle the dilemma). Finally, the right hand side of \cref{fig:tradeoffs} is dedicated to the technical enablers, with a correspondence to the specific trade-off challenges they can solve. Part of these technical enablers will be described in \cref{sec:tech_enablers}, with specific reference to the corresponding trade-offs and a figure summarizing their interactions (\cref{fig:paper_flow} in the sequel).} 
\subsection{First order effect: promoting sustainable 6G innovation}
With the ongoing exponential traffic growth, sustainable 6G networks must significantly reduce their \textit{negative first order effects}, striving toward the ultimate goal of achieving a \textit{Net Zero network}. This refers to a \SI{90}{\percent} reduction in \gls{ghg} emissions, with the remaining 10\% being offset through carbon removal efforts \cite{SBTi24}. \update{Major operators have committed to achieve this target between 2040 and 2050 (e.g., Orange by 2040~\cite{Orange2040}). Consequently, 6G must not only meet the challenge of extremely low energy consumption with a majority use of renewable energies, but also that of extreme sustainability in terms of its lifespan.} This can be achieved, for example, through \textit{minimizing energy consumption} during manufacturing, deployment, and operation, \textit{optimizing material usage} and ensuring \textit{recyclability}, as well as designing networks capable of withstanding increasingly \textit{challenging climate conditions}. Besides the environmental perspective, this must be accomplished while maintaining economic viability within the competitive landscape of the communication industry. 
This focus extends beyond communication to include other domains such as computation, control, actuation and learning. 
From a social perspective, this will ultimately foster greater acceptance of new technologies, as sustainability continues to rise in importance among both users and society as a whole.

\subsection{Second order effect: an overview and the case of 3 verticals}\label{sec:2ordereff}
\begin{figure*}  
\centering
\includegraphics[width=0.98\textwidth]{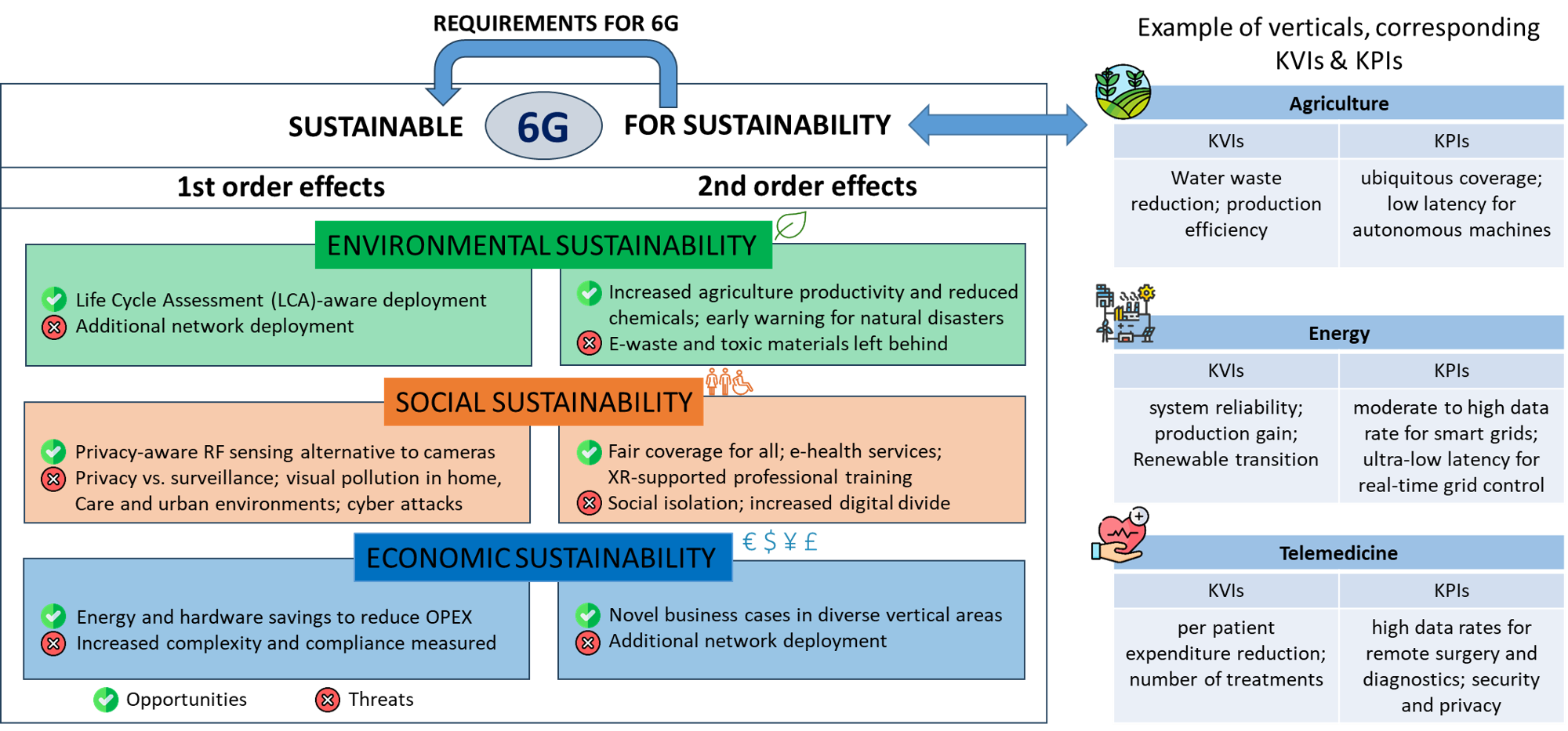}
    \caption{\update{Mapping of 1st and 2nd order effects, sustainability pillars, sustainability aspects and verticals with KVIs and KPIs.}}
\label{fig:6G_KVIs}
%\vspace{-0.35 cm}
\end{figure*}

\textit{6GS} aims to drive the \textit{positive second order effect} by improving the sustainability impacts from the use of 6G by different users including vertical sectors, through transformative initiatives. Each vertical sector possesses unique sustainability needs across the three pillars, often with differing priorities and emphases on specific goals. As an example, three vertical scenarios (\textit{Agriculture, Energy, and Telemedicine}) are elaborated in \cref{fig:6G_KVIs}. With a broader view, other usage scenarios across a broad array of vertical sectors demand a comprehensive range of communication technology capabilities, as outlined in the IMT-2030 framework \cite{IMT2030}. These include, among others, coverage, throughput, reliability, security, energy, and low latency. These capabilities collectively enable addressing diverse goals and values across the sustainability pillars. Beyond these examples, smart cities, transportation, education and manufacturing, have their own unique demands towards sustainability.
\begin{tcolorbox}[colframe=blue!10, colback=blue!10, coltitle=red, coltext=black]
A holistic perspective on sustainability impact pathways across the entire 6G ecosystem is needed \cite{HFF+21}.
\end{tcolorbox}
\section{Unfolding the dilemma: the key trade-offs}\label{sec:all_tradeoffs}
The overarching aspects introduced in \cref{sec:dilemma} (cf. \cref{fig:6Gwheel}) relate to values as much as they do to technological performance. \update{First, as recently summarized in \cite{SUSTAIN6G2025}, "\textit{Key Value} refers to principles or qualities that individuals or groups deem important, desirable, or intrinsically good that may be addressed or impacted by \gls{ict}." KVs can be also interpreted as criteria that drive innovation \cite{WIKSTROM2024}. However, a measurable metric is needed for the KVs to align outcome and impact against objectives \cite{SUSTAIN6G2025}. These measurable indicators are known as KVIs, and are usually measurable by, e.g., tracking progress in the context of a vertical use case \cite{WIKSTROM2024}. At the same time, KPIs still represent a valid way of assessing technology solutions.} It becomes clear that: (i) both \glspl{kvi} and \glspl{kpi} should be driving \gls{6g}; (ii) it is left open how trade-offs among supporting new services and overarching values could be made. 
\begin{tcolorbox}[colframe=blue!10, colback=blue!10, coltitle=red, coltext=black]
The dilemma introduced in \cref{sec:dilemma} can be analyzed via a set of trade-offs along KVIs and KPIs.
\end{tcolorbox}
This section introduces \gls{kvi}-related and \gls{kpi}-related trade-offs, illustrating the complexity inherent in navigating sustainability in communication technologies.
\subsection{KVI-related trade-offs across the three sustainability pillars}\label{sec:tradeoff}
Mobile connectivity is a critical resource to support people in crises and to promote their well-being. The advent of 6G hence raises social, economical, ethical, and geopolitical questions: Can technological innovation be driven at the same time by conventional KPIs and new KVIs? Would these counteract each other, or could they be mutual reinforcing? If difficult 'apples versus pears' trade-offs have to be made, are there any methodologies that could help to bridge the gaps? 

Profitability of industrial processes, retail, and logistics can increase through real-time and location-specific information. 
Precision farming enabled by connected sensors can optimize crops and minimize the use of fertilizers and other chemicals. But at which cost? It is evident that the multi-faceted targets lead to dilemmas: resilient networks will require some redundancy to avoid single points of failure, while ecological concerns would steer to lean network deployments avoiding redundancy; sensing technologies can enhance safety, yet they raise privacy concerns; \gls{ai} opens great opportunities to enhance services and network operation, yet comes with a significant energy consumption impact and potentially requires corrections to ensure fairness; 'connecting the unconnected' may enhance inclusiveness, yet economically serving densely populated and prospering areas is most beneficial; Is the necessity for a 6G technology today clear and sufficient to justify the negative impacts it will have? \update{ \cref{fig:6G_KVIs} introduces opportunities and threats for first and second order effects across all pillars. Exemplary KVIs and corresponding KPIs from agriculture, energy and telemedicine are provided, presenting requirements to S6G from the 6GS perspective to be considered in making trade-offs.}
\subsection{KPI-related trade-offs across technologies}
We now dive into more technical trade-offs covering different networking aspects towards solving the \textit{isolation of domains} issue (cf. \cref{sec:intro}) \update{and breaking down the overarching dilemma described in \cref{sec:dilemma}, and reduce it to technical questions. This overall flow is further detailed in \cref{sec:tech_enablers}.}

\subsubsection{\textbf{Availability vs. resource utilization}}\label{sec:res_vs_availability}
6G promises higher levels of service availability compared to previous generations. 
However, these are typically achieved through redundancy and additional tools that help to quickly identify failures and apply mitigation measures. These measures often induce more energy-intensive operations and require redundant hardware, leading to increased electronic waste at \gls{eol}. Hence, mechanisms are needed to increase utilization, e.g., by implementing dynamic resource allocation, and reduce \gls{eol} waste, by introducing modular designs to allow for easier repairs. Robust and fault-tolerant networks often require additional resources, such as back-up systems.
\subsubsection{\textbf{Communication vs. computing}}\label{sec:comm_vs_comp}
Focusing on in-network \gls{ai} (ubiquitous intelligence- cf. Section \ref{sec:tradeoff}) and its \textit{dichotomic role} of enabling and enabled technology, \gls{ai} is now used to optimize networks from the physical layer to network orchestration \cite{MerluzziHexa23}. 
However, while \gls{ai} and computing come with the benefit of enablers for improved network and vertical usage scenarios optimization, they also represent new sources of resource and energy consumption, for training and inference phases that take place constantly during  operation. 
A \textit{multidimensional trade-off} arises, comprising hardware resources such as graphical, tensor and neural processing units (GPUs, TPUs, NPUs), absolute energy consumption (for communication and computing aspects), accuracy, latency, privacy and trustworthiness. As a general observation, energy consumption must be monitored in a holistic fashion (cf. \cref{sec:joint_comm_comp}).  Although we cannot ignore and shall make the best use of the power of \gls{ai}, we must work to \textit{constantly monitor and optimize the values and associated costs in a closed loop}, including communication and computation aspects. 
\subsubsection{\textbf{Energy efficiency  vs. absolute energy consumption}}\label{sec:energy_vs_total_resource} 
Energy efficiency of transmission, expressed in transmit energy per bit, has progressed spectacularly. Each new network generation has improved this parameter by an order of magnitude or more~\cite{Sabella2016}. At the same time, the absolute energy consumption of mobile networks has increased. This can be attributed to the even much steeper increase in traffic, up to two orders of magnitude over a decade~\cite{EricssonMobilityReport}, caused by many more users and higher quality services. This illustrates the rebound effect: efficiency gains lead to higher overall resource usage due to increased demand. There is a clear need for metrics that comprehensively capture the entire spectrum of resource usage.
They must consider the full energy lifecycle and broader ecological first order effect (cf. \cref{sec:res_vs_availability}), including the consumption of rare earth materials and clean water throughout equipment manufacturing, deployment, operation, and disposal phases. To summarize: 
%\vspace{-0.1 cm}
\begin{tcolorbox}[colframe=blue!10, colback=blue!10, coltitle=red, coltext=black]
\begin{itemize}
    \item The values generated by availability increase the risk of waste across the full life cycle;
    \item Bit-level efficiency metrics fall short when both communication and computation are involved;
    \item Improving energy efficiency does not necessarily translate into energy consumption reduction
\end{itemize}
\end{tcolorbox}
%\vspace{-0.3 cm}
\section{Technology enablers}\label{sec:tech_enablers}
\begin{figure}[t]
    \centering
    \includegraphics[width=0.98\columnwidth]{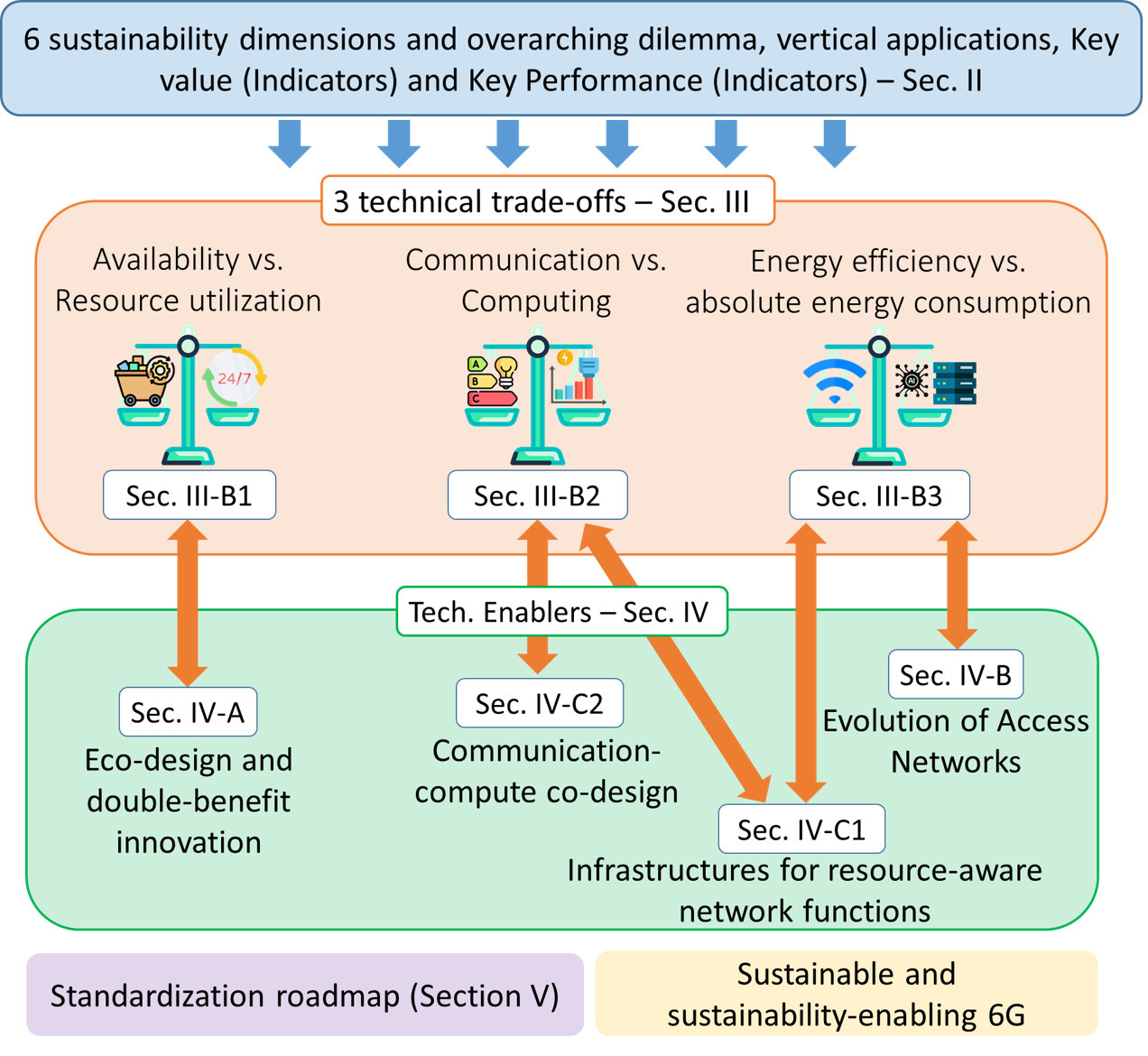}
    \caption{\update{Breaking down the dilemma into trade-offs and solutions: the view of this work. Link between dimensions, trade-offs and a subset of technology enablers}}
    \label{fig:paper_flow}
    %\vspace{-0.3 cm}
\end{figure}
We now present a subset of the technology enablers (the full spectrum is illustrated in \cref{fig:tradeoffs}) that we believe will drive 6G towards striking the best balance among the introduced trade-offs.  
\update{\cref{fig:paper_flow} summarizes the methodology flow of this work, which breaks down the overarching dilemma across the six sustainability dimensions into the set of trade-offs, and finally to the technical solutions with the corresponding sections.} 
\update{\subsection{Eco-design and double-benefit innovation}\label{sec:eco_design}}
\update{Design has conventionally focused on performance versus cost of equipment and operation. As introduced in \cref{sec:res_vs_availability}, this approach is too wasteful in terms of energy, water, and materials, even if cost factors call for designing low-power solutions. A paradigm shift towards eco-design aims to reduce the ecological impact of 6G networks and devices, by integrating environmental aspects into the product development process. Eco-design should consider all phases of the life cycle  (production, operation, transportation, and end-of-life).
To date, electronic design often results in highly integrated solutions and "sealed" packaging, which are difficult or impossible to repair or disassemble. A modular eco-design concept prioritises repairability and recycling from the outset.}

\update{We advocate to balance
environmental and economic requirements where needed, yet strive for "double-benefit" innovation that can make both go hand-in-hand. The following example clarifies such an opportunity. Networked-controlled repeaters, proposed as novel networking elements for 6G, offer performance upgrades at low complexity~\cite{10908557}. Requiring much less power than conventional base stations, they do not need cooling, having lower impact in production and operation. If implemented in a form factor small enough to be transported as small parcel by public mail services, also transportation and EoL could benefit from lower cost and ecological impact.}\\
\update{Last but not least, policies are essential to promote sustainable approaches in electronics design. Recent legal frameworks introducing the “right to
repair” are an important step forward.}
\subsection{Evolution of Access Networks}
\label{sec:RAN}
\update{To address the trade-off introduced in \cref{sec:energy_vs_total_resource}, energy efficiency must overcome the increase in traffic load. Much of this should happen in the \gls{ran}, which remains the most energy-consuming segment of wireless networks. While many of the efforts in the evolution of the \gls{ran} so far have focused on providing quantitative performance improvements in terms of mobile network KPIs, the current trajectory of services and applications does not indicate a need for a significant performance leap per user. Indeed, while 5G performance requirements were ambitious and remain relevant for many scenarios, extreme values will only be needed for highly specific applications with dedicated deployments, and most performance indicators for future networks will remain in the same order of magnitude as current ones, aligning with the lower range of estimated objectives set by ITU IMT-2030 research. Instead, we argue that the evolution of RAN must prioritize cost reduction, environmental sustainability, and absolute energy consumption. This will enable new, profitable use cases and services while minimizing environmental impact. Besides achieving a net reduction of the total energy consumed by the mobile network, the design of future networks should also address operational and social requirements, including minimal end-to-end environmental impact, high resilience, digital inclusion, security, and exposure awareness.}
\\
In parallel, the evolution of optical networks plays a crucial role in the sustainability of 6G. All-Optical Networks (AONs) offer photonic continuity in converged fixed-wireless scenarios. AON has the potential to lower energy consumption, eliminate electromagnetic fields, and 80\% reuse across generations and substantial reductions in energy consumption \cite{SinghFiber20}. 
\subsection{Resource-aware network management beyond RAN}\label{sec:comp_management}
As previously mentioned in Sections \ref{sec:energy_vs_total_resource} and \ref{sec:comm_vs_comp}, the \textit{per-bit} metrics driving energy-aware wireless networking for the last decade, are becoming limited, also due to the increasingly prominent role of in-network computing and \gls{ai}. 
\subsubsection{\textbf{Infrastructures for resource-aware network functions}}\label{sec:infrast_comp}
As current trends on virtualization and centralization bring much of the mobile network protocol stack into the (edge and central) cloud, computing plays a central role in the mobile network and the energy consumed by computational tasks represents a significant burden \cite{cloudric}. In this context, the sustainable operation of the cloud infrastructure becomes critical to address the trade-off introduced in \cref{sec:energy_vs_total_resource}. Traditional approaches to virtualize the protocol stack rely on Hardware Accelerators (HAs), which ensure a very high performance yet also consume much more energy as compared to other computing platforms such as traditional CPUs (e.g., the energy consumption of typical GPU-based HA can be more than one order of magnitude higher than that of a traditional CPU \cite{cloudric}). Further, in typical deployments there is one HA co-located in each base station, which is highly inefficient in terms of network deployment (raising sustainability issues regarding the use of materials and creation of electric waste).\\
Computational resources must be allocated to achieve an optimal trade-off between energy consumption and performance. 
Architectures must also enable dynamic sharing of the resources among multiple tasks, which is challenging today. Finally, the network protocol stack needs to be improved to handle variations in compute availability, ensuring reliable task completion and graceful degradation during outages.
\subsubsection{\textbf{Communication-compute co-design}}\label{sec:joint_comm_comp}
\update{In \cref{sec:comm_vs_comp}, we introduced the importance and associated challenges that the role of computing introduces in future wireless networks.} It becomes then clear that, although independently optimizing \gls{ran} (\cref{sec:RAN}) and computing aspects (\cref{sec:infrast_comp}) can bring sustainability gains, this convergence makes a joint management and orchestration the most promising way to achieve the holistic perspective, and to explore the trade-offs described in \cref{sec:comm_vs_comp} at best~\cite{MerluzziHexa23}. 
On one hand, the advancement of communication technologies can help accessing computing resources in unprecedentedly fast and reliable fashion. On the other hand, ubiquitous computing can help opportunistically extracting relevant information (semantics), to reduce redundancy that does not serve the specific scope of communication (goal-oriented communication). 
However, as highlighted in \cref{sec:all_tradeoffs}, this requires a balance of resources from both aspects, thus calling for their joint design and an exchange of information, to generate a complete view on energy and resource consumption at both segments. 
\update{This is of course a challenging task. While today there is no energy information exchange between these different segments, exposure functionalities exist in, e.g., ETSI specifications to retrieve radio network information at Multi-Access Edge Computing (MEC) system side \cite{ETSI_MEC}. Extending this type of functionalities should enable a lean information sharing between different systems, thus having access to end-to-end measures to realize a joint approach}. 
Going further, the power of \gls{ai} extensively demonstrated its potential in optimizing network at different layers, from physical to resource orchestration in multi-agent systems, \update{answering some questions introduced in \cref{sec:res_vs_availability}}. Today, \gls{ai} is more and more offered \textit{as a service}~\cite{MerluzziHexa23}. In this direction, moving computing towards the network edge (i.e., closer to the data source and the end service consumers) is a promising solution to reduce backbone network load, enhanced privacy and reduce latency, thus potentially contributing to social values. This leads to the concept of \textit{edge AI}, whose scope is embed intelligence in resource limited devices with respect to central clouds \cite{Mao24}, with benefits in terms of latency and privacy, as data are kept as local as possible.
\subsection{\update{What does the holistic perspective need to become reality?}}\label{sec:smp}
All listed innovations (and those shown in \cref{fig:tradeoffs}) are fundamental bricks for a holistic view of 6G, and to explore all introduced trade-offs at best. \update{Once mapped to use cases with sector-specific sustainability needs, alternative technical solutions will typically arise, with different balance between first and second order effects. Therefore, fundamental questions arise on how to favour and select a solution over others, knowing that stakeholder interests and sustainability needs typically differ. How to resolve potentially conflicting goals in terms of performance and sustainability across sectors? It becomes then clear that a negotiation or a recommendation mechanism should take place, to weight sustainability needs for the 6G system, against 6G network performance towards sustainability needs of vertical sectors. This negotiation requires all involved stakeholder to share monitored data (or proxy metrics to avoid exposure of potentially sensitive information) and priorities. Creating such a \textit{\gls{smp}} demands the definition of new interfaces and protocols for this negotiation, data format and temporal aspects, as well as security and privacy as core values to be guaranteed. 
Not only technical, but also political and market-related facets play a role. Research and consequent regulation efforts must be spent in this direction, to realize this complex yet fundamental interaction able to balance first and second order effects.} 
\section{Towards implementing sustainability with 6G}\label{sec:standardization}
A roadmap towards 6G standardization with sustainability principles should define concrete actions, engage stakeholders, and align innovation with global goals, across \gls{3gpp} and other complementary organizations such as O-RAN Alliance (for cloud-native infrastructure) and TM Forum (for open APIs).  
Since 5G networks, enhanced with \gls{ai} and cloud technologies, can meet foreseeable service needs, there is no urgency for a disruptive 6G shift,  
but rather for impactful, deployable features that address clear business needs while considering all sustainability dimensions. 
\update{At European level, one of the core objectives of the SUSTAIN-6G project is to define the above mentioned strategic roadmap. 
As part of a collective effort from operators to vendors and vertical use case owners, contributions to various \glspl{sdo} are being prepared, as described in what follows.
Within \gls{3gpp}, and in particular RAN1, major operators such as Orange focus on the topics of coding, MIMO, sensing, and Agentic AI, always with a sustainability perspective. Proposals on LDPC (R1-2506148) featuring new optimized matrices that reduce processing time and computational complexity are planned. Additionally, proposal R1-2506394, entitled “Views on 6G PHY choices,” aims to minimize the costs associated with \gls{6g} deployment through maximum hardware reuse, in line with \cref{sec:res_vs_availability} and \cref{sec:eco_design} discussion. Further, particularly in SA5, an ongoing Work Item addresses topics aligned with those covered in ETSI EE and ITU-T SG5.}

\update{Sustainability considerations are being emphasized in RAN, especially within RAN3, to ensure their inclusion in the forthcoming 6G system from Release 21 onwards. Within ETSI EE, the operator Telecom Italia (TIM) serves as the Rapporteur for the specification on RANs metrics and methods of measurement of energy efficiency (ES 203 228), which has also been adopted by the European Commission Code of Conduct. In ITU-T, these ETSI specifications are reflected in the corresponding recommendations under SG5 Q6/5 (L.1331).} 

\update{Another fundamental aspect is related to use cases. As an example, vendors such as Ericsson are preparing proposals for use cases related to Massive IoT (M-IoT) and for defining new KPIs within the agreed use case for the utility sector. The control of future power grids will impose stricter requirements on bit rate and latency compared to existing IoT technologies; thus, this contribution aims to introduce new KPIs covering latency, payload size, and transfer intervals. In SA2, Ericsson contributes to topics such as policy control for user equipment energy requirements, user equipment energy consumption notifications, and energy optimization mechanisms for user equipment. Additionally, ITU-T SG5 Q9 focuses on methodologies for environmental impact assessments of ICT systems, including the development of simplified life cycle assessment approaches and updates to the sectoral footprint methodology. In ETSI EE, Ericsson plans to contribute to the development of energy measurement methods for servers and energy benchmarking tools (cf. \cref{sec:comm_vs_comp}).} 
\begin{tcolorbox}[colframe=blue!10, colback=blue!10, coltitle=red, coltext=black] 
Today, the activity towards the definition of a new standard has just started, and the moment is deemed to be the most appropriate to influence the definition of the standard with the inclusion of sustainability goals. 
\end{tcolorbox}

\section{Conclusions}\label{sec:conclusions}
We presented a view on dilemmas, trade-offs and technical innovations that will drive the design of 6G with sustainability as a core horizontal value from environmental, social, and economic perspectives. Starting from the overarching need of balancing first and second order effects, we identified trade-offs that go beyond legacy concepts leveraging energy efficiency. We believe that these trade-offs will drive technological innovations, to achieve the challenging objective of a sustainable by design 6G, enabling sustainable applications. The road ahead is still long and the biggest mountain to climb, beyond \gls{kvi}-driven technical challenges, is represented by \glspl{sdo}, to make eco-innovation at the center of 6G standardization, and to address the complex balance between first and second order effects across diverse stakeholders' priorities.

\bibliographystyle{IEEEtran}
\bibliography{Main_arxiv}
\newpage
\begin{IEEEbiographynophoto}{Dr. Mattia Merluzzi} (Member, IEEE) is a senior research engineer and project manager in wireless communications at CEA-Leti, Grenoble, France. 
\end{IEEEbiographynophoto}
\begin{IEEEbiographynophoto}{Olivier Bouchet} is a research engineer and project manager at Orange Labs, Rennes, France. He is the SUSTAIN-6G Technical Manager.
\end{IEEEbiographynophoto}
\begin{IEEEbiographynophoto}{Dr. Ali Balador} 
is a senior researcher and project manager in wireless networks at Ericsson Research, Sweden. 
\end{IEEEbiographynophoto}
\begin{IEEEbiographynophoto}{Prof. Gilles~Callebaut} (Member, IEEE) 
is a professor at KU Leuven, Belgium, working in the research group Dramco. 
\end{IEEEbiographynophoto}%
\begin{IEEEbiographynophoto}{Anastasius Gavras} is a programme and project manager in telecommunications at Eurescom GmbH in Heidelberg, Germany 
\end{IEEEbiographynophoto}%
\begin{IEEEbiographynophoto}{Prof. Liesbet~Van~der~Perre} (Senior Member, IEEE) is a professor at KU Leuven, Belgium, and a guest professor at ULund, Sweden. 
\end{IEEEbiographynophoto}%
\begin{IEEEbiographynophoto}{Prof. Albert Banchs} (Senior Member, IEEE) is a professor at University Carlos III and director of IMDEA Networks institute, Madrid, Spain.
\end{IEEEbiographynophoto}
\begin{IEEEbiographynophoto}{Mauro Renato Boldi} is currently with TIM, where he is in charge of European projects in the context of networks, cloud and edge, sustainability.
\end{IEEEbiographynophoto}
\begin{IEEEbiographynophoto}{Dr. Emilio Calvanese Strinati} (Member, IEEE) is a research engineer and project manager in wireless communications at CEA-Leti, Grenoble, France.
\end{IEEEbiographynophoto}
\begin{IEEEbiographynophoto}{Dr. Bahare M. Khorsandi} is Senior Staff Research Engineer at Nokia, Germany. She is coordinator deputy and WP lead in SUSTAIN-6G project.
\end{IEEEbiographynophoto}
\begin{IEEEbiographynophoto}{Dr. PhD. Marja Matinmikko-Blue} is Research Director at University of Oulu and Director of Sustainability and Regulation in 6G Flagship.
\end{IEEEbiographynophoto}
\begin{IEEEbiographynophoto}{Lars Christoph Schmelz} is Principal Research Lead at Nokia, Germany, and coordinator of the EU Sustainability Lighthouse project SUSTAIN-6G.
\end{IEEEbiographynophoto}

\end{document}